\documentclass[graybox]{svmult}

\usepackage{mathptmx}       
\usepackage{helvet}         
\usepackage{courier}        
\usepackage{type1cm}        
\usepackage{makeidx}         
\usepackage{graphicx}        
\usepackage{multicol}        
\usepackage[bottom]{footmisc}

\makeindex             

\begin{document}

\title*{New Insights into X-ray Binaries}
\author{Jorge Casares}
\institute{J. Casares \at Instituto de Astrof\'\i{}sica de Canarias, E-38200 La
Laguna, Tenerife, Spain, \email{jcv@iac.es}}
%
%
\maketitle


\abstract{X-ray binaries are excellent laboratories to study collapsed
objects. On the one hand, transient X-ray binaries contain the best examples of 
stellar-mass black holes while persistent X-ray binaries mostly
harbour accreting neutron stars. The determination of stellar masses  
in persistent X-ray binaries is usually hampered by the overwhelming luminosity 
of the X-ray heated accretion disc. However, the discovery of high-excitation
emission lines from the irradiated companion star has opened new routes 
in the study of compact objects. This paper presents novel techniques 
which exploits these irradiated lines and summarises the dynamical masses 
obtained for the two populations of collapsed stars: neutron stars and black 
holes.}

\section{Introduction}
\label{sec:1}
X-ray binaries (XRBs hereafter) are interacting binaries where X-rays arise 
from the accretion of matter onto a neutron star (NS) or a black hole (BH).  
Accretion processes are found in other astrophysical environments such as 
cataclysmic variables (i.e. interacting binaries with accreting white dwarfs), 
T Tauri stars, protoplanetary discs, etc. but the unique  property of XRBs 
is the presence of central compact objects that are the remnants of collapsed,  
massive stars. Therefore, they provide the best 
laboratories to study their properties in detail, such as masses, spin or 
NS equation of state.  
This paper is not meant to give a thorough review of XRBs but 
instead it will focus on three selected topics with implications  
for our knowledge of the mass spectrum of collapsed stars. These are: 

\begin{enumerate}
\item{{\bf Evidence for BHs in XRBs:} a summary of dynamical masses is 
presented.}
\item{{\bf The Bowen Project:} a new technique to trace the orbit of 
companion (donor) stars in persistent XRBs.}
\item{{\bf Echo Tomography:} reprocessed light from the donor star can constrain
the binary inclination, the parameter with largest impact in the mass error.} 
\end{enumerate}

Excellent reviews on other aspects of XRBs can be found in ~\cite{char06} 
and ~\cite{mcc06}. 
   
\section{Black Holes in X-ray Binaries}
\label{sec:2}
The mass distribution of Black Holes (BHs hereafter) has 
strong impact in several areas of Astrophysics, in particular SNe models, 
the evolution of massive stars, chemical enrichment of the galaxy, jet 
formation etc.. Stellar evolution theories predict $\sim10^9$ BHs remnants in 
the Galaxy~\cite{brown94} but only BHs in interacting binaries can be easily 
detected through X-ray radiation triggered by accretion. 
This is the reason why the history of BH discoveries has run in parallel with 
the development of X-ray astronomy. 

\begin{figure}[ht]
\includegraphics[width=0.6\textwidth]{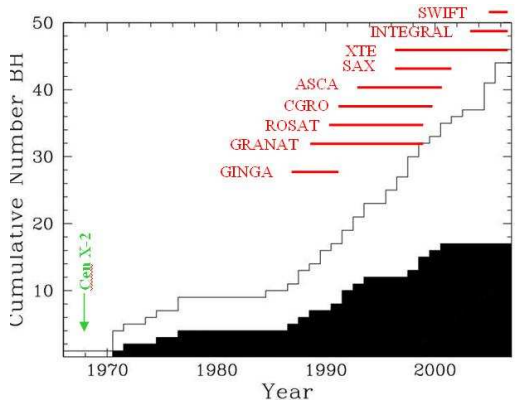}
\hfill
\parbox[b]{0.39\textwidth}{\caption{Cumulative distribution of BH SXTs 
discovered during the era of X-ray
astronomy. The black histogram indicates dynamical BHs \newline}}
\label{fig:1}
\end{figure}

In particular Soft X-ray transients (SXTs hereafter) provide the best 
systems to find stellar-mass BHs, since $\sim$75\% of these transients 
likely harbour a BH. SXTs are a subclass of X-ray binaries with low-mass donor 
stars (typically of K-M spectral types) which exhibit episodic outbursts due to 
mass transfer instabilities in the accretion disc \cite{king99}. 
During outburst, the X-ray luminosity increases by factors up to 
10$^6$-10$^7$ and, therefore, they are easily spotted by X-ray satellites. 
Unfortunately, the companion star is overwhelmed by the X-ray heated disc 
at all wavelengths, precluding its detection.  
However, after a few months of activity, X-rays switch off, the reprocessed 
light drops off several magnitudes into quiescence and  the companion starts 
to dominate the optical flux. This provides a special opportunity for 
spectroscopic detection, perform dynamical studies and derive stellar masses. 
Figure~1 shows the cumulative histogram of BH SXTs discovered 
since 1966, when Cen X-2 was first detected during a rocket
flight. A linear increase at a rate of $\sim$1.7 yr$^{-1}$ is apparent since 
the late 80's, when a new fleet of X-ray satellites with 
higher sensitivity and All-Sky-Monitor capabilities became operative.

The best evidence for the presence of BHs is dynamical, i.e. a 
compact object whose mass exceeds the maximum allowed for stable neutron 
stars or $\sim$3 M$_{\odot}$ ~\cite{rho74}. And this is relatively easy to 
prove through the observed radial velocity curve of the companion star during 
quiescence. The orbital period $P_{\rm orb}$ and velocity semi-amplitude 
$K_{\rm C}$ combine in the standard mass function equation $f(M)$   
\begin{eqnarray}
f(M) = {P_{\rm orb} K_{\rm C}^3\over 2 \pi G} = {M_{\rm X} \sin^3 i\over
(1+q)^2}~~~~~~ ,\nonumber
\end{eqnarray}
which relates the mass of the compact object $M_{\rm X}$ and the companion 
star $M_{\rm C}$ through the orbital inclination $i$ and mass ratio 
$q={M_{\rm C}\over M_{\rm X}}$. It is easy to show that $f(M)$ yields a 
secure lower  limit to the BH mass $M_{\rm X}$.  
This experiment typically requires resolving powers $\ge$1500 and can be
performed with targets brighter than $R\sim$23 using current instruments on
10m-class telescopes. 

But getting actual BH masses rather than lower limits is not so straightforward. 
The reason being that, by their very nature, BHs do not burst nor pulse and 
hence one cannot trace their orbital motion. We are facing a single-line 
spectroscopic binary where the extra observables ($q$ and $i$) must be 
extracted from the optical star. And this can be accomplished with two further 
experiments:

\begin{itemize}

\item{{\it Resolving the rotational broadening} of 
the donor's photospheric lines. Because the donor star is filling 
the Roche lobe and synchonized, the rotational broadening $V \sin i$ scales 
with $K_{\rm C}$ as a function of $q$~\cite{wade88}. Therefore, by measuring $V \sin i$ one 
can determine $q$ directly.}

\item{{\it Fitting synthetic models to the 
ellipsoidal modulation}. The changing visibility of the tidally distorted 
companion star generates a double-humped light curve (the so-called  
ellipsoidal modulation) whose amplitude is a strong function of $i$ and $q$. 
For extreme mass ratios $q<0.2$, typical of BH binaries, the shape of
the light curve is weakly sensitive to $q$ and hence $i$ can be easily 
determined~\cite{sha94}.} 
\end{itemize} 

By combining the mass function with constraints on $q$ and $i$ one gets a full 
dynamical solution and hence the BH mass with minimum assumptions. Further
details on this prescription and possible systematics involved can be found in
several reviews such as ~\cite{casa01}.  

BHs have also been found in a handful of High Mass X-ray Binaries (HMXBs 
hereafter), i.e. XRBs with early-type massive donor stars. However, here we 
find several limitations which complicate the analysis. A key factor is 
$M_{\rm C}$, which for a HMXB is a large number and has a wide 
range of uncertainty. The optical star is likely to be undermassive for its 
spectral type as a result of mass transfer and binary evolution~\cite{rap83}. 
Furthermore, 
mass transfer is usually produced through winds rather than Roche lobe 
overflow and this has a two-fold effect. On one side, wind emission can 
contaminate the radial velocities of the donor star. On the other, since 
the optical star does not fill its Roche lobe, $q$ and $i$ 
values derived through $V \sin i$ and ellipsoidal fits may be
overestimated. These caveats can only be side-lined in eclipsing 
binaries such as X-7 in M33, which so far is the only case where this has 
been possible. 
In addition to the eclipse duration, the distance provides an extra 
restriction which lead to tight constraints in the space parameter. 
In particular, the radius of 
the donor, the Roche lobe filling factor and the inclination are 
accurately determined and yield a very precisse BH mass ~\cite{oro07}.

\begin{table}
\caption{Dynamical BHs}
\label{tab:1}       
%
%
\begin{tabular}{lccccc}
\hline\noalign{\smallskip}
 System &  $P_{\rm{}orb}$ &  $f(M)$ & Donor  &  Classification & $M_{\rm x}$~$^{\dagger}$ \\
 &   [days] &  [$M_{\odot}$] &  Spect. Type & &  [$M_{\odot}$] \\ 
\noalign{\smallskip}\svhline\noalign{\smallskip}
GRS 1915+105$^a$    &     33.5    &    9.5 $\pm$ 3.0        &    K/M III   & LMXB/Transient   &   14 $\pm$ 4  \\
V404 Cyg        &      6.471  &   6.09 $\pm$ 0.04       &    K0 IV     &      ,,          &   12 $\pm$ 2   \\
Cyg X-1         &      5.600  &  0.244 $\pm$ 0.005      &    09.7 Iab  &  HMXB/Persistent &   10 $\pm$ 3   \\
LMC X-1$^b$         &      3.909  &   0.143 $\pm$ 0.007       &    07 III    &     ,,          &  10.3 $\pm$ 1.3          \\
M33 X-7         &      3.453  &   0.46 $\pm$ 0.08       &    07-8 III  &      ,,          &   15.7 $\pm$ 1.5 \\
XTE J1819-254   &      2.816  &   3.13 $\pm$ 0.13       &    B9 III    &  IMXB/Transient  &  7.1 $\pm$ 0.3 \\ 
GRO J1655-40    &      2.620  &   2.73 $\pm$ 0.09       &    F3/5 IV   &      ,,          &  6.3 $\pm$ 0.3 \\
BW Cir$^c$ &    2.545  &   5.73 $\pm$ 0.29       &    G5 IV     &  LMXB/Transient  &    $>$ 7.0     \\	 
GX 339-4$^d$        &      1.754  &   5.8  $\pm$ 0.5        &     --       &      ,,          &   $>$ 6.0     \\
LMC X-3         &      1.704  &   2.3  $\pm$ 0.3        &    B3 V      &  HMXB/Persistent &  7.6 $\pm$ 1.3 \\
XTE J1550-564   &      1.542  &   6.86 $\pm$ 0.71       &    G8/K8 IV  &  LMXB/Transient  &  9.6 $\pm$ 1.2 \\
4U 1543-475     &      1.125  &   0.25 $\pm$ 0.01       &    A2 V      &  IMXB/Transient  &  9.4 $\pm$ 1.0 \\
H1705-250       &      0.520  &   4.86 $\pm$ 0.13       &    K3/7 V    &  LMXB/Transient  &    6 $\pm$ 2   \\
GS 1124-684     &      0.433  &   3.01 $\pm$ 0.15       &    K3/5 V    &      ,,          &  7.0 $\pm$ 0.6 \\
XTE J1859+226$^e$  &  0.382  &   7.4  $\pm$ 1.1        &     --       &      ,,          &                \\
GS2000+250      &      0.345  &   5.01 $\pm$ 0.12       &    K3/7 V    &      ,,          &  7.5 $\pm$ 0.3 \\
A0620-003       &      0.325  &   2.72 $\pm$ 0.06       &    K4 V      &      ,,          &   11 $\pm$ 2   \\
XTE J1650-500   &      0.321  &   2.73 $\pm$ 0.56       &    K4 V      &      ,,          &                \\
GRS 1009-45     &      0.283  &   3.17 $\pm$ 0.12       &    K7/M0 V   &      ,,          &  5.2 $\pm$ 0.6 \\
GRO J0422+32    &      0.212  &   1.19 $\pm$ 0.02       &    M2 V      &      ,,          &    4 $\pm$ 1   \\
XTE J1118+480   &      0.171  &   6.3  $\pm$ 0.2        &    K5/M0 V   &      ,,          &  6.8 $\pm$ 0.4 \\
\noalign{\smallskip}\hline\noalign{\smallskip}
\end{tabular}

$^{\dagger}$ Masses compiled by \cite{char06} and \cite{oro03}. \\
$^a$ New photometric period of 30.8$\pm$0.2 days reported by 
\cite{neil06}.\\ 
$^b$ Updated after \cite{oro08}. \\
$^c$ Updated after \cite{casa08}. \\
$^d$ Updated after \cite{munoz08}. \\
$^e$ Period is uncertain. See \cite{zur02}.   
\end{table}

Table~\ref{tab:1} presents an updated list of confirmed BHs based on 
dynamical arguments, with their best mass estimates. We currently have 
21 BHs, with orbital periods between 33.5 days and 4.1 hours. 
The great majority are SXTs (17) while 4 are persistent HMXBs: 
Cyg X-1 plus the 3 extragalactic binaries LMC X-1, LMC  X-3 
and M33 X-7. 
The case of GX 339-4 deserves special mention because it is the only SXT 
where the presence of a BH was proven during the outburst phase. This was 
possible thanks to the detection of fluorescent lines arising from the
irradiated companion (see Sect.~\ref{sec:3}). GRS 1915+105 is also noteworthy, 
not only because of its long orbital period and large mass function but also 
because IR spectroscopy was essential to overcome the $>25$ magnitudes of 
optical extinction and reveal the radial velocity curve of the 
companion star ~\cite{grei01}. However, it should be noted that recent
photometry reports a slightly shorter orbital period and evidence for 
irradiated light curves ~\cite{neil06}. The combination of these two 
effects will likely decrease the mass function and BH mass. XTE J1859+226 
also needs revisiting because its orbital period is uncertain ~\cite{zur02}. 
In summary we have 16 BH masses  ranging between 4 and 16 M$_{\odot}$ 
with $\sim$5-30 \% errors.
These can be compared with theoretical distributions of stellar remnants 
such as ~\cite{fryer01}. The model includes 
binary interaction under Case C mass transfer (i.e. Common Envelope evolution 
after core helium ignition), wind mass-loss in the Wolf-Rayet phase and SN Ib 
explosion. The computation predicts a continuum distribution of remnants with 
a mass cut at 12 M$_{\odot}$ which is difficult to reconcile with some 
of the observed masses. 
However, the model entails many theoretical uncertainties 
which dominate the final mass spectrum such as the Common Envelope 
efficiency, the wind mass-loss rate or the progenitor's mass cut. 
Clearly more SXT discoveries and lower uncertainties in BH masses  
are required before these issues can be addressed and the
form of the distribution is used to constrain BH formation models and 
XRB evolution. 
 
\begin{figure}[ht]
\includegraphics[width=0.45\textwidth, angle=-90]{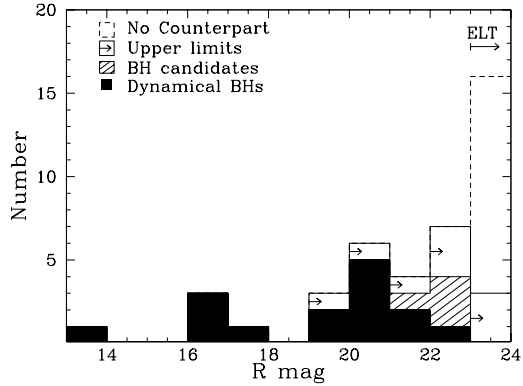}
\hfill\par\vspace*{-9em}\hfill
\parbox[b]{0.39\textwidth}{\caption{Magnitude distribution of BH SXTs in
quiescence. 
The black histogram indicates dynamical BHs while the rest are BH 
candidates. Targets without optical counterpart are likely fainter 
than $R>23$. \newline}}
\label{fig:2}
\vspace*{3em}
\end{figure}

In addition to dynamical BHs, there are 27 other SXTs with similar X-ray
spectral and timing properties during outburst\footnote{This number has been 
updated after \cite{mcc06} with new detections reported in several Astronomical
Telegrams.}. 
Unfortunately, these BH {\it candidates} become too faint in quiescence for 
dynamical studies or even lack accurate astrometry. 
This is illustrated in 
Fig.~2 which shows the magnitude distribution of the 44 currently known 
BH transients. Dynamical studies are only possible with the current largest 
telescopes for sources brighter than $R\le23$. Not shown in the 
figure is the heavily reddened GRO 1915+105 which was studied in the NIR. 
The figure depicts the bright tail of a {\it dormant}  
population of galactic BH SXTs which several works have estimated in a few 
thousand systems (\cite{romani98} and included references). Improving the 
statistics of dynamical BHs requires not only  a new generation of ELT 
telescopes to tackle fainter targets but also new strategies aimed at 
unveiling new ``hibernant" SXTs before they go into outburst. 
Quiescent BH SXts typically have  EW (H$_{\alpha}$)$\sim20-50$ \AA~ and hence 
they should show up in deep H$_{\alpha}$ surveys such as IPHAS~\cite{drew05}. 
However, clever diagnostics need
to be defined to clear out other populations of H$_{\alpha}$ emitters such as 
cataclysmic variables or T Tauris (see Corral-Santana et al., these 
proceedings). 

\section{The Bowen Project}
\label{sec:3}

Aside from transient XRBs, there are $\sim$150 persistent XRBs in the Galaxy, 
the great majority hosting neutron stars (NS hereafter) accreting at 
the Eddington limit. 
They are considered the progenitors of Binary Millisecond Pulsars (BMPs 
hereafter) because is the sustained accretion during their long active lives 
that spins the NS up to millisecond periods. The discovery of millisecond 
pulses in 8 
transient XRBs and coherent oscillations during X-ray bursts in 13 
persistent XRBs gave strong support to this ``recycle" pulsar scenario. 
And burst oscillations were detected in addition 
to persistent pulses in the transient XRBs SAX J1808-3658 \cite{chak03} 
and XTE J1814-338 \cite{stro03} with identical frequencies. 
This confirmed that burst oscillations are indeed modulated with the spin 
of the NS. The interest of these discoveries stands in the fact that one can 
use the orbital Doppler shift of pulses/oscillations to trace the NS orbit and 
obtain the X-ray mass function. 

Optical emission in persistent XRBs is triggered by reprocessing of the 
intense X-ray radiation in different binary sites, mainly the accretion disc. 
The companion star is $\sim$1000 times fainter than 
the irradiated disc at optical-IR wavelengths and hence completely undetectable. 
This has systematically plagued attempts to determine system parameters and, 
in most cases, only the orbital period is known. Fortunately, there are methods 
which can exploit the effects of irradiation and X-ray variability.
New prospects were opened by the discovery of sharp high excitation 
emission lines arising from the irradiated face of the companion star in 
Sco X-1~\cite{stee02}. The most prominent are found in the core of the Bowen 
feature, a blend of CIII/NIII lines which are mainly powered by fluorescence. 
These lines trace the motion of the companion star and 
provided the first dynamical information on this protypical LMXB (see 
Fig.~3). Since then, sharp Bowen lines 
from companion stars have been discovered in 7 other persistent LMXBs and 
4 transients during outburst: Aql X-1, GX 339-4 and the BMPs XTE J1814-338 
and SAX J1808.4-3658. These transient studies 
beautifully demonstrate the power of this technique in systems which otherwise 
cannot be studied in quiescence because either they  are too faint (case of 
GX 339-4 and the BMPs) or are contaminated by a bright interloper (Aql X-1). 
In particular, the case of GX 339-4 is remarkable because the Bowen study 
provides the first solid evidence for the presence of a BH in this classic 
transient. 

\begin{figure}[ht]
\begin{center}
\begin{picture}(250,190)(50,30)
\put(0,0){\includegraphics{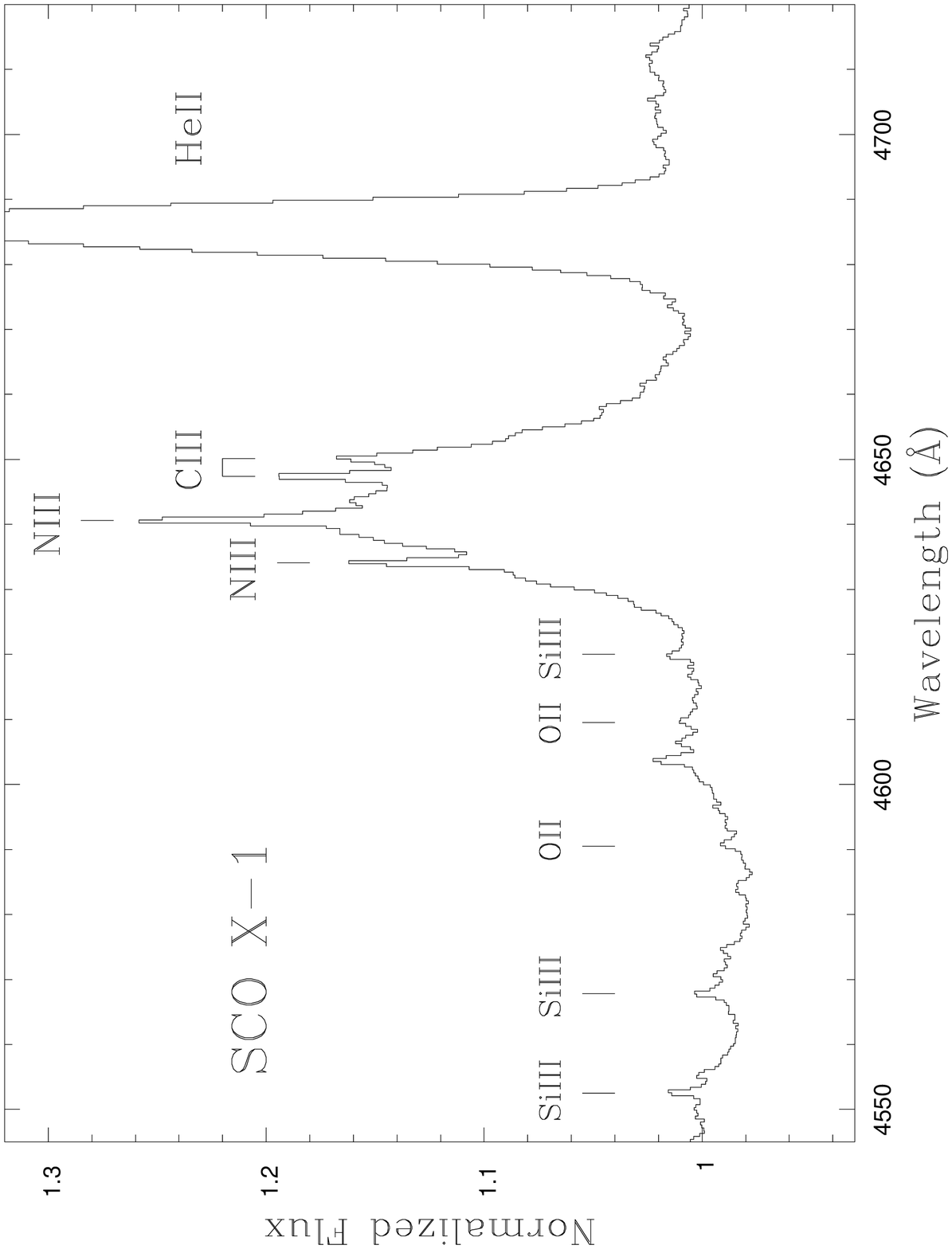}}
\put(0,0){\includegraphics{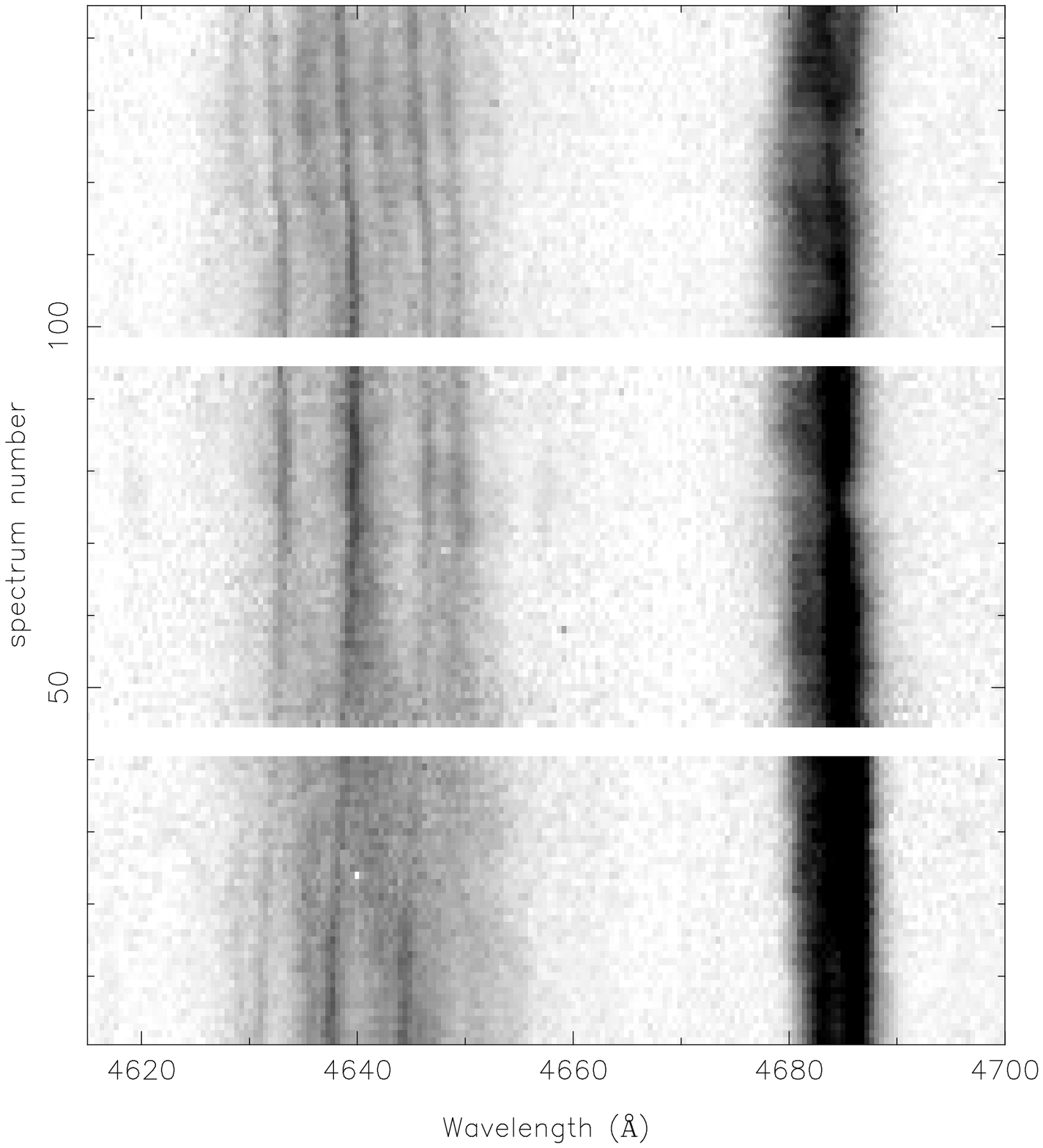}}
\noindent
\end{picture}
\end{center}
{\caption{Detecting companion stars in persistent XRBs. Left: the main 
high excitation emission lines due to irradiation of 
the donor star in Sco X-1. Adapted from \cite{stee02}. 
Right: Trail spectrum showing the radial velocity motion of the Bowen CIII/NIII lines 
as a function of time.   
After \cite{stee02}.}} 
\label{fig:3}
\end{figure}

The radial velocity curves of the Bowen lines are biased because they arise 
from the irradiated face of the star instead of its center of mass. Therefore, 
a {\it K-correction} 
needs to be applied in order to obtain the true velocity semi-amplitude $K_{\rm C}$ 
from the observed velocity $K_{\rm em}$. 
The {\it K-correction} parametrizes the displacement of the center of light with 
respect to the donor's center of mass through the mass ratio and disc 
flaring angle $\alpha$. The latter dictates the size of the disc shadow 
projected over the irradiated donor~\cite{munoz05}.  
Extra information on $q$ and $\alpha$ is thus required to get the real 
$K_{\rm C}$ . Furthermore,  useful limits to the NS mass can be set if the binary 
inclination is well constrained through eclipses.

\begin{table}
\caption{NS masses obtained using the Bowen Technique}
\label{tab:2}       
%
%
\begin{tabular}{lllllll}
\hline\noalign{\smallskip}
 System &  $P_{\rm{}orb}$ &~~~~~  Mag &~~~~~ Type  &~~~~~  $K_{\rm em}$ &~~~~~ $M_{\rm
 NS}$~$^{\dagger}$ & Reference\\
 &   [hr] &   &  &~~~~~ [km/s] &~~~~~  [$M_{\odot}$] &\\ 
\noalign{\smallskip}\svhline\noalign{\smallskip}
Sco   X-1            & 18.9&~~~~~  B=12.2    &~~~~~  Persistent&~~~~~  87  $\pm$ 1   &~~~~~ $>$0.2     &~\cite{stee02} \\
LMC   X-2            & 8.1 &~~~~~  B=18    &~~~~~  ,,          &~~~~~  351 $\pm$ 28  &~~~~~  $>$ 1.2 &~\cite{corne07c} \\
X1822-371            & 5.6 &~~~~~  B=15.8  &~~~~~   ,,         &~~~~~  300 $\pm$ 15  &~~~~~   1.6-2.3&~\cite{casa03} \\
V926 Sco (X1735-444) & 4.7 &~~~~~  B=17.9  & ~~~~~  ,,         &~~~~~  226 $\pm$ 22  &~~~~~   $>$ 0.5&~\cite{casa06} \\
GX9+9  (X1728-16)    & 4.2 &~~~~~  B=16.8  & ~~~~~  ,,         &~~~~~  230 $\pm$ 35  &~~~~~   $>$ 0.3&~\cite{corne07b} \\
GR Mus               & 3.9 &~~~~~ B=19.1  & ~~~~~  ,,          &~~~~~  245 $\pm$ 30  &~~~~~  1.2-2.6 &~\cite{barnes07}\\
V801 Ara (X1636-536) & 3.8 &~~~~~  B=18.2  & ~~~~~  ,,         &~~~~~  277 $\pm$ 22  &~~~~~   $>$ 0.8&~\cite{casa06} \\ 
EXO 0748-676         & 3.8 &~~~~~  B=16.9  &~~~~~   ,,         &~~~~~  310 $\pm$ 10  &~~~~~   1.1-2.6&~\cite{munoz09} \\
\noalign{\smallskip}
\hline\noalign{\smallskip}
Aql X-1              & 19 &~~~~~  V $\sim$ 22 &~~~~~ Transient &~~~~~  247 $\pm$ 8 &~~~~~  $>$1.6&~\cite{corne07a} \\
GX 339-4             & 42.1 &~~~~~ V $>$21  &~~~~~    ,,       &~~~~~ 317 $\pm$ 10 &~~~~~  $>$6.0&~\cite{hynes03}\cite{munoz08} \\
XTE J1814-338$^a$        & 4.2 &~~~~~  V= 23.3 & ~~~~~ Transient/BMP    &~~~~~  345 $\pm$ 19  &~~~~~ $>$1.0   &~\cite{casa09} \\ 
SAX J1808.4-3658         & 2.0 &~~~~~  V= 21.8 & ~~~~~   ,,     &~~~~~  248 $\pm$ 20  &~~~~~ $>$0.3   &~\cite{corne09} \\ 
\noalign{\smallskip}\hline\noalign{\smallskip}
\end{tabular}

$^{\dagger}$ After K-correction and constraints to the inclination, mass ratio 
or NS velocity (when available).\\
$^a$ Preliminary results. \\   
\end{table}

Table ~\ref{tab:2} summarises the NS masses obtained through the Bowen 
technique during several campaigns at the WHT, AAT and VLT. The list of 
persistent systems is almost a complete sample of Galactic LMXBs brigther 
than B$\simeq$19. 
In the cases of Aql X-1 and X1822-371 the evidence of NS more massive 
than canonical is very persuasive. The latter is a particularly favourable 
binary because it is eclipsing and the NS is a pulsar. Then its radial velocity 
curve is known through the study of orbital pulse delays. Good constraints on 
the NS velocity are also available for V801 Ara through the detection of pulse 
oscillations during a superburst~\cite{casa06}. Tight limits to the inclination 
and mass ratio are also available for the eclipsing EXO 0748-676~\cite{munoz09} 
and the dipper GR Mus~\cite{barnes07}. 
In the remaining cases the NS mass is not well constrained due to large 
uncertainties in the inclination and/or mass ratio. 
However, it is important to stress that these are the first dynamical 
constraints in persistent LMXBs since their discovery, 40 years ago. 
Other techniques (such as the Echo Tomography) need to be exploited to 
further refine these limits and derive more accurate NS masses.   
Previous reviews presenting results of the Bowen project can be found in 
\cite{casa04} and \cite{corne08}. 

\section{Echo Tomography}
\label{sec:4}
Echo Tomography uses time delays between X-ray and UV/optical variability as a 
function of orbital phase to map the reprocessing sites in a binary~\cite{obrien02}. 
The optical variability can be modelled by the convolution of the X-ray 
light curve with a {\it transfer function} which depends on the binary geometry. 
The {\it transfer function} encodes information on the most fundamental 
parameters such as the binary inclination, star separation and mass ratio. And 
in particular, the component associated with the companion star is most 
sensitive to these parameters so detecting echoed emission from the donor  
offers the best opportunity to constrain them. There has been several attempts 
at detecting correlated optical and X-ray variability using white light or broad 
band filters (e.g.~\cite{van90},~\cite{hynes05}). These works have detected delays 
which are mostly consistent with reprocessing in the outer disc implying that 
the disc is the dominant source of continuum reprocessed light.

\begin{figure}[ht]
\begin{center}
\begin{picture}(250,190)(50,30)
\put(0,0){\includegraphics{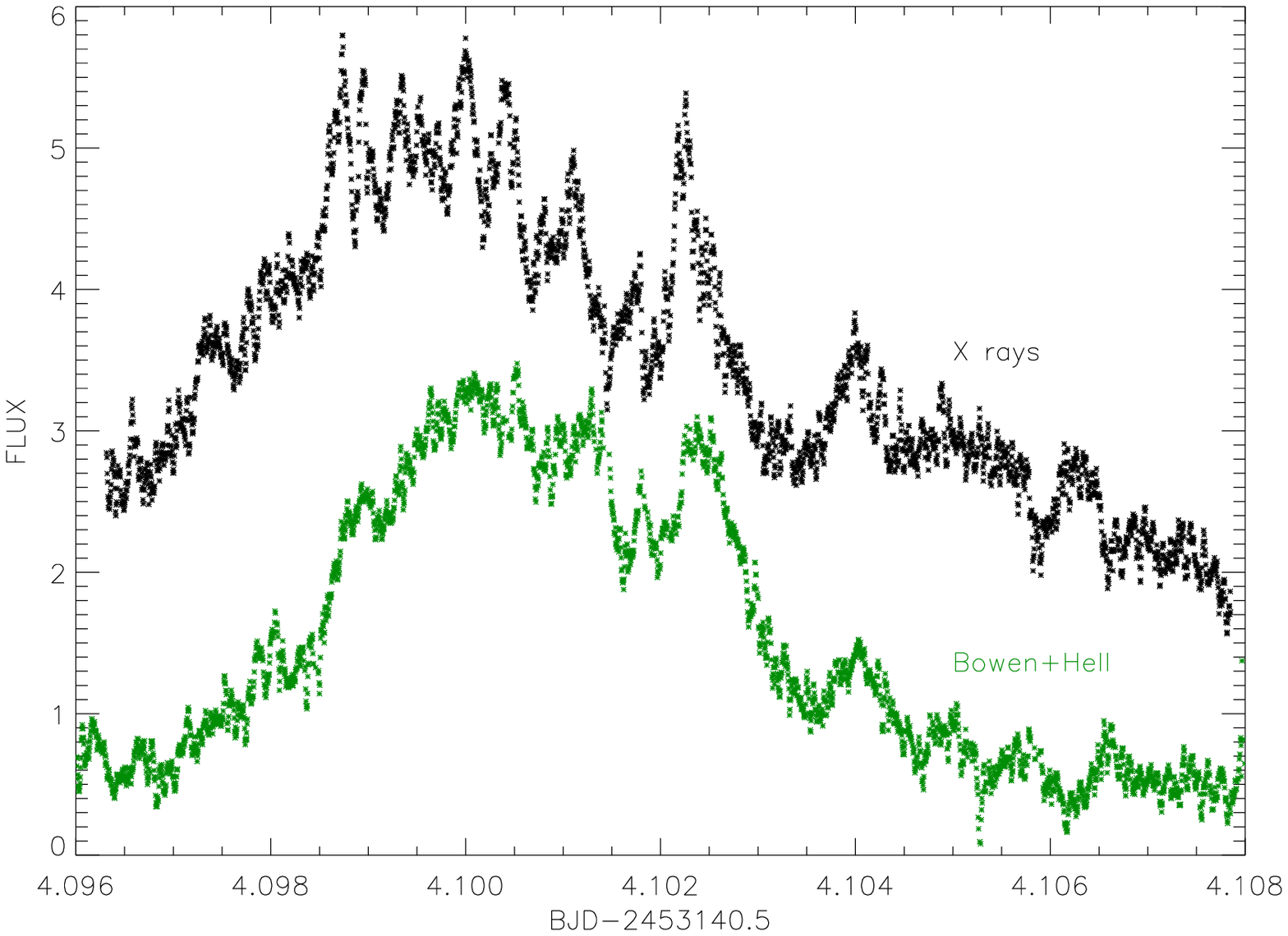}}
\put(0,0){\includegraphics{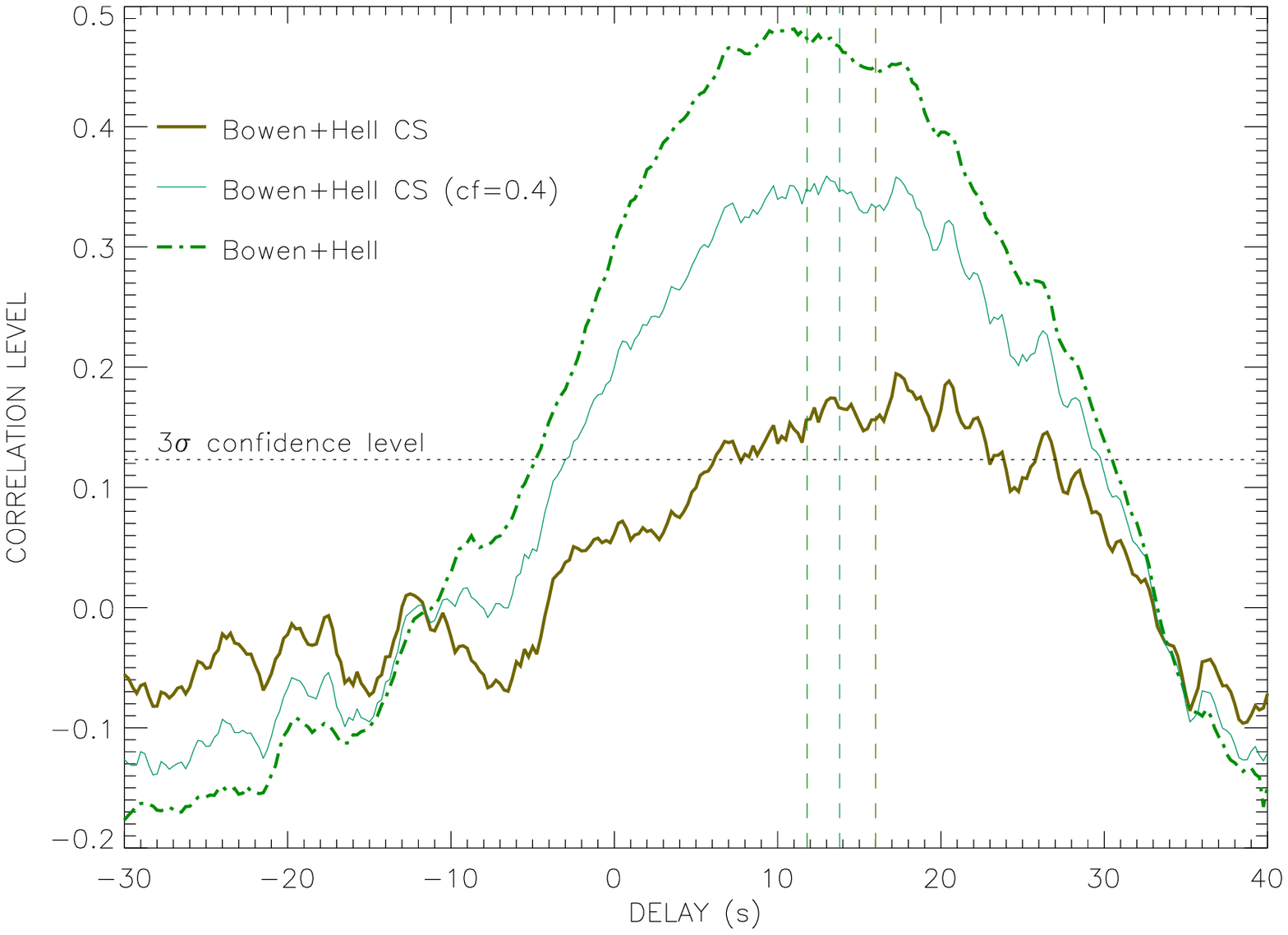}}
\noindent
\end{picture}
\end{center}
{\caption{Echo Tomography experiment in Sco X-1. Left: large amplitud X-ray 
variability and correlated optical light curve observed at orbital phase 0.5. 
Sco X-1 happened to be in the flaring branch state.   
Right: Cross-correlation functions between X-rays and optical light curves 
observed in the  continuum (top), Bowen+HeII window (middle) and  Bowen+HeII 
after continnum subtraction (bottom). 
After~\cite{munoz07}.}} 
\label{fig:4}
\end{figure}

Exploiting emission-line reprocessing rather than 
broad-band photometry has two potential benefits: a) it amplifies the response 
of the donor's contribution by suppressing most of the background
continuum light (dominated by the disc); b) since the
emission line reprocessing time is instantaneous, the response is
sharper (i.e. only smeared by geometry). 
Through the Bowen project we know that high energy radiation is very 
efficiently reprocessed by the donor's atmospheres into Bowen 
fluorescence lines. 
Therefore, we decided to search for optical echoes of X-ray variability using 
ULTRACAM~\cite{dhillon07} equipped with a special set of narrow band filters, 
centered at the Bowen blend and a red continuum. The latter is essential to 
subtract the continuum light and hence amplify the reprocessed signal from the 
companion. During an RXTE/WHT campaign on Sco X-1 correlated variability was 
detected at phase $\simeq0.5$ i.e. superior conjunction of the companion star, 
when the heated face presents its maximum visibility~\cite{munoz07}. Time 
delays of 14-16s are measured after the continuum light is subtracted from the 
Bowen light curves (see Fig.~4). These delays are consistent with 
the light traveltime between the NS and the companion star and hence provide 
the first evidence of reprocessing in the companion of Sco X-1. However, one 
needs to detect several optical echoes as a function of orbital phase in order 
to constrain $i$ and $q$ and derive masses.  

\begin{figure}[ht]
\begin{center}
\begin{picture}(250,190)(50,30)
\put(0,0){\includegraphics{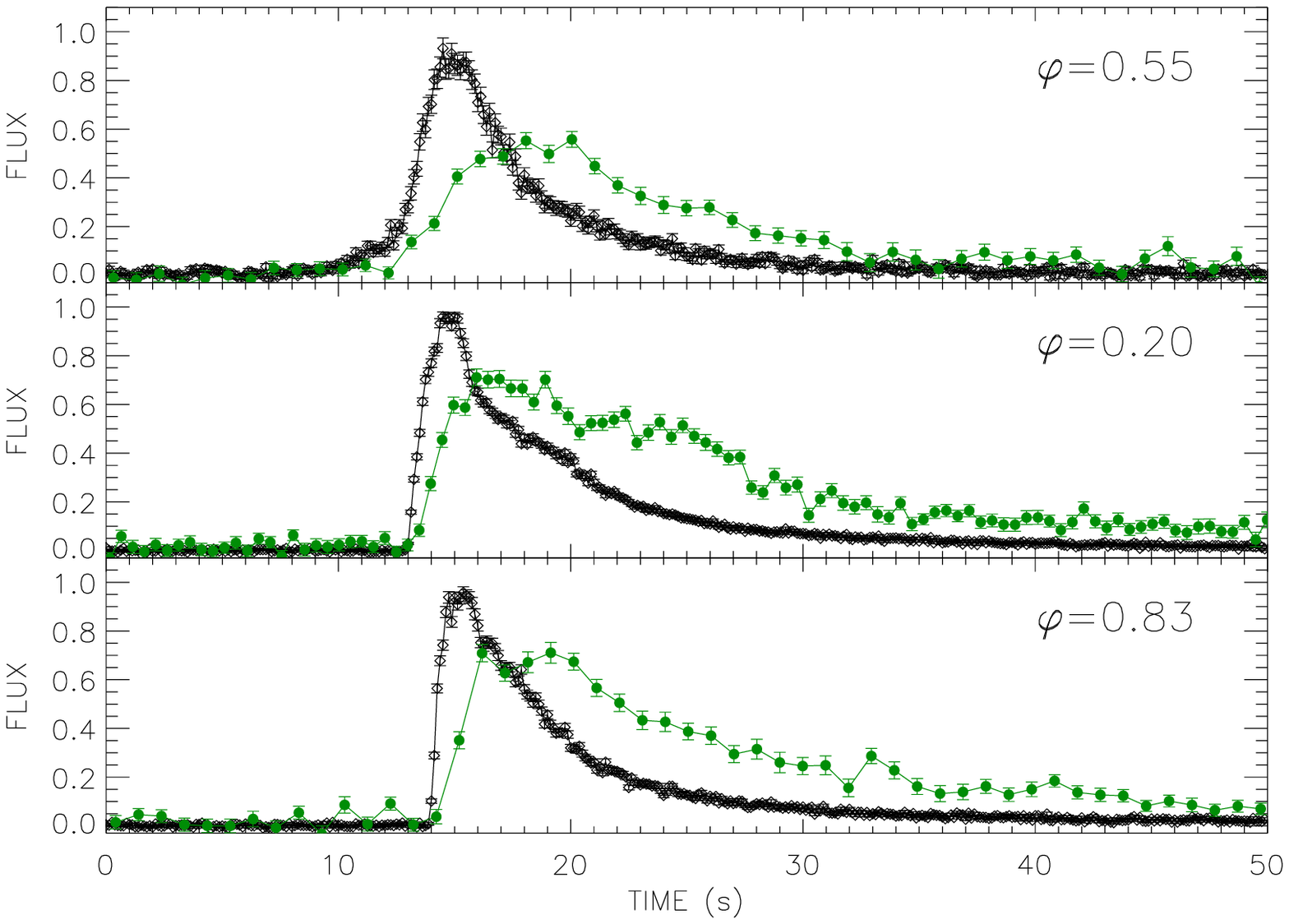}}
\put(0,0){\includegraphics{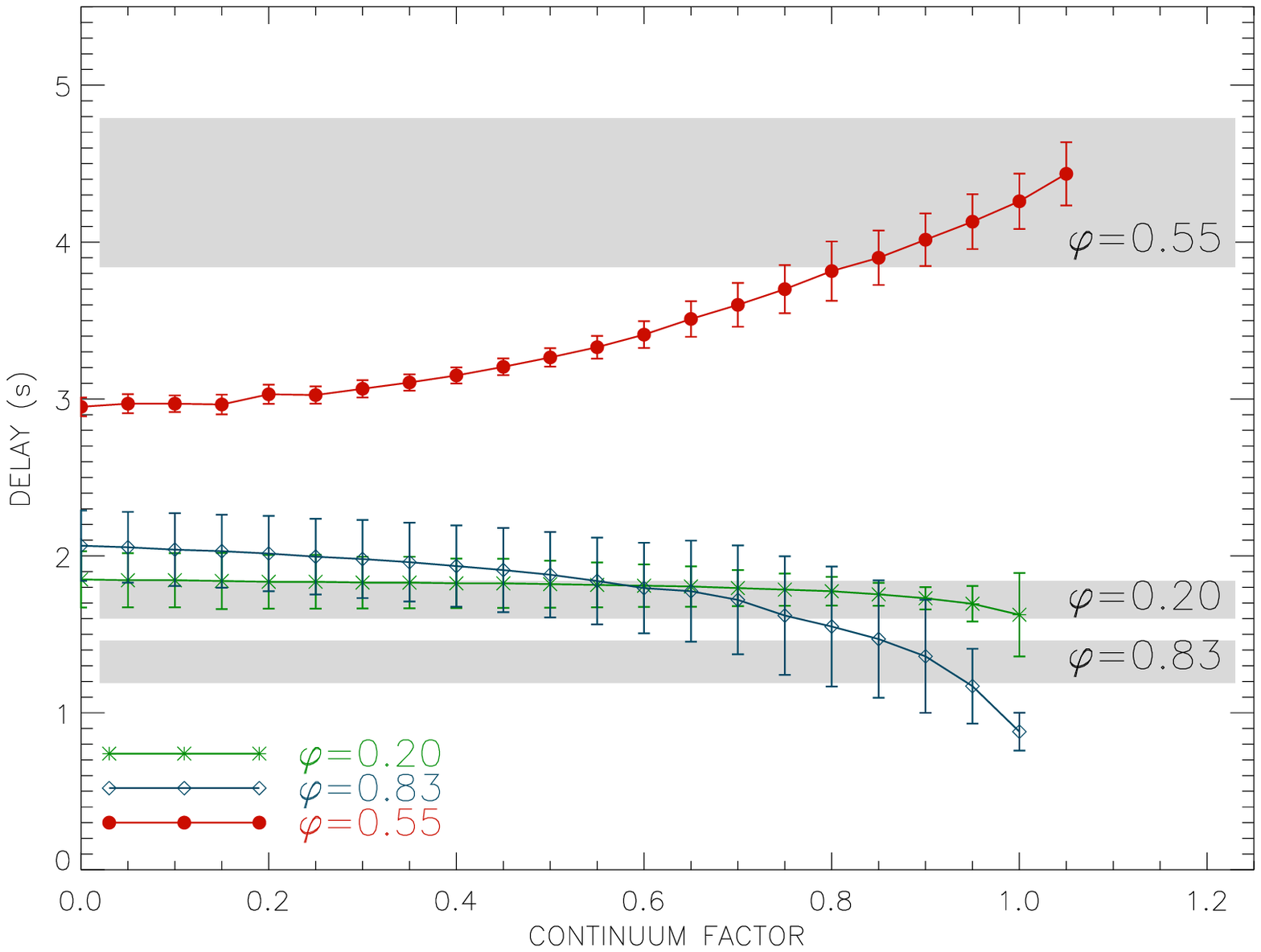}}
\noindent
\end{picture}
\end{center}
{\caption{Echo Tomography of X1636-536. Left: the three X-ray bursts detected 
and their optical (Bowen+HeII) counterparts. 
Right: delay times between the X-ray and Bowen+HeII burst light curves as a function 
of continuum subtraction factor. Shaded regions correspond to delays expected for 
reprocessing in the companion at each orbital phase. They are computed for M$_{\rm NS}=1.4$ 
M$_{\odot}$, $q=0.3$, $i=36-60^{\circ}$ and $\alpha=12^{\circ}$. 
After ~\cite{munoz08b}.}} 
\label{fig:5}
\end{figure}
 
In a second campaign we observed the burster X1636-536 simultaneously with 
RXTE and VLT+ULTRACAM. 
Three X-ray bursts and their corresponding optical echoes were recorded 
at orbital phases 0.55, 0.20 and 0.83 and these are shown in the left panel of 
Fig.~5. The optical bursts clearly lag X-ray burst and are also 
smeared, indicating an extended reprocessing site. 
Delay times are in the range 2-3 secs showing little evidence for orbital 
variability. However, these delays drift when several amounts of 
continuum light (parametrised by the factor {\it cf}) are subtracted from the
Bowen+HeII light. And for $cf\simeq 0.8-0.95$ the 3 delays become consistent 
with reprocessing in the companion for  M$_{\rm NS}=1.4$  
M$_{\odot}$, $q=0.3$, $\alpha=12^{\circ}$ and $i=36-60^{\circ}$ , as derived
through radial velocities of the Bowen lines ~\cite{casa06}. This is illustrated
in the right panel of Fig.~5. Note that, in particular, delays 
observed at phase $\sim0.5$ are especially sensitive to the inclination angle. 
The main difficulty which hinders us from constraining the inclination is 
the unknown amount of continuum substraction.   
In principle, there must be an optimum {\it cf} factor which results in a perfect 
subtraction.  
However, this is not easy to find because 
the continuum filter is placed $\sim$1500 \AA~ away from the Bowen lines due 
to the optical layout of ULTRACAM. 
	New high-speed spectrophotometry devices such as ULTRASPEC will provide  
	pure emission line light curves for echo mapping experiments.
	These are likely to yield accurate inclinations and, when combined 
	with dynamical information from the Bowen lines and X-ray mass
	functions, the first accurate NS masses in persistent XRBs.

\section{Conclusions}
\label{sec:5}

\begin{svgraybox}
In the past 20 years the field of X-ray binaries has experienced 
significant progress with the discovery of 17 new BHs and 8 transient 
BMPs in LMXBs. Dynamical masses are available for 16 BHs but better 
statistics and improved errors are required before using the observed 
distribution to constrain XRB evolution and supernova models. Exploiting deep 
H$\alpha$ surveys of the Galactic plane, such as IPHAS, may unveil 
a significant fraction of a large expected population of quiescent XRBs. 

The discovery of fluorescence emission from the companion star has opened 
the door to derive NS masses in persistent and new transient XRBs. 
This is possible thanks to: i) dynamical information from irradiated 
donors through high-resolution spectroscopy
of the Bowen blend; ii) echo-mapping reprocessing sites through
simultaneous Bowen-line/X-ray lightcurves. These techniques, together
with results from burst oscillations and transient BMPs, 
will likely provide the first accurate NS masses in XRBs 
in the near future and perhaps confirm the existence of massive 
NS. 
Thanks to these new techniques, which have proven their worth, the future 
is bright as new instruments and telescopes will allow to push ahead our 
sample of BHs and NS masses.  
High-speed and high-resolution instruments, such as OSIRIS at GTC, RSS at 
SALT and ULTRASPEC, will play a crucial role in this goal.
\end{svgraybox}

\begin{acknowledgement}
I would like to acknowledge helpful comments from my colleagues D. Steeghs, 
R. Cornelisse and T. Mu\~noz-Darias. I'm also grateful for support 
from the Spanish MEC grant AYA2007-66887. Partially funded by the Spanish 
MEC under the Consolider-Ingenio 2010 Program grant CSD2006-00070: First Science 
with the GTC. 
\end{acknowledgement}
%

%
%
%
\biblstarthook{}

\end{document}